\documentclass[a4paper,11pt]{article}
\usepackage{graphicx,amssymb,amstext,amsmath}
\usepackage{color}

\usepackage{hyperref}
\hypersetup{
   bookmarks=true,         
   unicode=false,          
   pdftoolbar=true,        
   pdfmenubar=true,        
   pdffitwindow=false,     
   pdfstartview={FitH},    
   pdftitle={General Twist Fields},    
   pdfauthor={James Silk},     
   pdfsubject={Twist Fields},   
   pdfcreator={James Silk},   
   pdfproducer={}, 
   pdfkeywords={}, 
   pdfnewwindow=true,      
   colorlinks=true,       
   linkcolor=blue,          
   citecolor=red,        
   filecolor=magenta,      
   urlcolor=cyan           
}

\newcommand{\bc}{\begin{center}}
\newcommand{\ec}{\end{center}}
\def\ba#1{\begin{array}{#1}\displaystyle}
\newcommand{\ea}{\end{array}}
\newcommand{\z}{\\[2mm] \displaystyle}

\newcommand{\beq}{\begin{equation}}
\newcommand{\eeq}{\end{equation}}
\newcommand{\beqa}{\begin{eqnarray}}
\newcommand{\eeqa}{\end{eqnarray}}
\newcommand{\no}{\nonumber}
\newcommand{\n}{\nonumber\\}
\newcommand{\bi}{\begin{itemize}}
\newcommand{\ei}{\end{itemize}}

\def\lt#1{\left#1}
\def\rt#1{\right#1}

\def\b#1{\bar{#1}}
\def\frc#1#2{\frac{#1}{#2}}

\newcommand{\p}{\partial}
\newcommand{\pr}{\partial_{r}}

\newcommand{\vac}{{\rm vac}}
\newcommand{\bra}{\langle}
\newcommand{\ket}{\rangle}

\newcommand{\rvac}{|\vac\rangle}
\newcommand{\Z}{{\mathbb{Z}}}

\newcommand{\R}{{\mathbb{R}}}

\newcommand{\T}{{\cal T}}
\newcommand{\D}{{\cal D}}

\renewcommand{\d}{{\rm d}}

\newcommand{\St}{\tilde{\Sigma}}
\newcommand{\psit}{\tilde{\psi}}

\date{\today}
\title{More General Correlation Functions of Twist Fields From Ward Identities in the Massive Dirac Theory}
\author{James Silk\footnote{Email: j.b.silk@durham.ac.uk} \\ Department of Mathematical Sciences, Durham University, Science Laboratories, \\ South Road, Durham, DH1 3LE, UK }

\begin{document}

\maketitle

\abstract{Following on from previous work we derive the non-linear differential equations of more general correlators of $U(1)$ twist fields in two-dimensional massive Dirac theory. Using the conserved charges of the double copy model equations parametrising the correlators of twist fields with arbitrary twist parameter are found. This method also gives a parametrisation of the correlation functions of general, fermionic, descendent twist fields. The equations parametrising correlators of primary twist fields are compared to those of the literature and evidence is presented to confirm that these equations represent the correct parametrisation.}

\tableofcontents

\section{Introduction}

Correlation functions of local fields of Quantum Field Theories contain all the physical information of the model. These functions are often only accessible perturbatively and at large energies, but in two-dimensional, integrable quantum field theories with free fermion representations non-trivial correlation functions can be expressed as solutions to integrable differential equations. The first example of this phenomenon is the spin-spin correlation function in the thermally perturbed Ising model \cite{WMTB75} and other results, some concerning models with Dirac fermions, have followed through a variety of methods: holonomic quantum fields (Dirac model) \cite{MSJ78}, Fredholm determinants (Ising and Dirac models) \cite{BB92,KBI93,BL97}, determinants of Dirac operators (Dirac model in flat and curved space and with magnetic field) \cite{P90,PBT94,L08}, and doubling of the model (Ising spin chain, and Ising QFT model at zero and non-zero temperature and in curved space) \cite{P80,FZ03,DF04}. The existence of such differential equations allows the relevant correlation functions to be evaluated very accurately once initial conditions have been fixed via conformal perturbation theory and form factor analysis.

Twist fields are local fields which have non-trivial monodromy, with respect to the fermion fields, associated to a symmetry of the model. From the holonomic quantum field description we expect general correlation functions of twist fields to be parametrised by integrable differential equations. Twist fields were first introduced in \cite{ST78}, as the $\Z_{2}$ monodromy fields of the Majorana fermion corresponding to the spin field of the Ising model. Similarly, it is known that primary $U(1)$ twist fields in the Dirac model reproduce correlation functions of exponential fields in the sine-Gordon model  at a particular value of the coupling (the free-fermion point) \cite{LZ96}. It is correlation functions of such $\Z_2$ and $U(1)$ primary twist fields that have been studied and that are known to lead to differential equations in QFT models.

This paper is an extension of the work presented in \cite{DS11} where the differential equations for the correlation function of twist fields with equal monodromy were found using a method similar to that of \cite{FZ03}. In \cite{FZ03} a double copy of Ising field theory is considered and its conserved charges used to derive Ward identities which lead to differential equations for the spin-spin two point function. The states considered in \cite{DS11} were kept as general as possible and only translation and parity symmetry were required in order to reduce the Ward identities to the known form, \cite{BL97}. In the present work we use the same methods to retrieve the differential equations parametrising the correlation functions of twist fields with differing monodromy. To obtain these equations it transpires that the states under consideration are also required to be rotation symmetric as without this property the Ward identities derived below do not provide sufficient information to obtain the result.

The twist fields of the Dirac model, $\sigma_{\alpha}$ for $\alpha\in\R$, are characterised by their $U(1)$ monodromy, $e^{2\pi i\alpha}$ and dimension $\alpha^{2}$. The properties of these fields, along with their fermionic descendents, are discussed in more detail in \cite{DS11}. The main aim of this work is to reproduce the differential equations parametrising the correlation function of twist fields. When written as
\begin{equation} \label{eq:defSig}
 \langle\sigma_{\alpha}(x,y)\sigma_{\beta}(0,0)\rangle= c_{\alpha}c_{\beta} m^{\alpha^{2}+\beta^{2}}e^{\Sigma(x,y)}
\end{equation}
where $\Sigma$ is a function of $r=(1/4)(x^{2}+y^{2})$ we find that it is the a solution of the following equations, fist found in \cite{SMJ78-3}:
\begin{align}
 \left(\pr^{2}+\frac{1}{r}\pr\right)\Sigma & =\frac{m^{2}}{2}(1-\cosh(2\psi)) \label{eq:result1} \\
 \left(\pr^{2}+\frac{1}{r}\pr\right)\psi & =\frac{m^{2}}{2}\sinh(2\psi)+\frac{(\alpha-\beta)^{2}}{r^{2}}\tanh\psi(1-\tanh^{2}\psi). \label{eq:result2}
\end{align}

In deriving these equations we find a discrepancy between (\ref{eq:result2}) and the results of \cite{BL97}, namely a factor of $4$ multiplying the $(\alpha-\beta)^{2}$ is not present in our equations. As we are unable to find an error, or explain this missing factor, we attempt to verify which equation is correct. Using the form factor expansion of the correlator at large $r$ we obtain an approximate expression for $\Sigma$ and thus also for $\psi$. We then examine how well this expression solves both (\ref{eq:result2}) and the equation of \cite{BL97}. Numerical and analytical methods are used to this end with each one indicating that the equations presented in this paper do indeed give the correct parametrisation.

The method presented in this paper also gives differential equations for the correlation functions of descendent twist fields,
\begin{align}
  \langle\sigma_{\alpha,\alpha+1}(x,y)\sigma_{\beta+1,\beta}(0,0)\rangle = & pe^{i\theta(\alpha-\beta)}c_{\alpha}c_{\beta}m^{\alpha^{2}+\alpha+\beta^{2}+\beta+1}e^{\Sigma'(r)} \no \\
  \langle\sigma_{\alpha+1,\alpha}(x,y)\sigma_{\beta,\beta+1}(0,0)\rangle = & qe^{i\theta(\beta-\alpha)}c_{\alpha}c_{\beta}m^{\alpha^{2}+\alpha+\beta^{2}+\beta+1}e^{\Sigma'(r)} 
\end{align}
parametrised by the same auxiliary function $\psi$. Both correlators are shown to have the same $r$ dependence and the constants $p$ and $q$ are discussed in more detail in section \ref{sec:symcorfunc}, but essentially depend on the positions of the fields through their braiding relations. It is shown the function $\Sigma'$ satisfies
\begin{equation}
  \left(\pr^{2}+\frac{1}{r}\pr\right)\Sigma' = \frac{(\pr\psi)^{2}}{\sinh^{2}\psi}-m^{2}-\frac{(\alpha-\beta)^{2}}{r^{2}\cosh^{2}\psi}
\end{equation}
which is a new result that appears naturally from the Ward identities.

This paper is set out as follows: In section \ref{sec:model} notation is set out and the model under consideration is explained. We also present the fields and relevant form factors, however no derivation is presented and we refer the interested reader to \cite{DS11} for more details. Section \ref{sec:DoubleMod} introduces the double model and the action of relevant conserved charges is described. Again more detail is presented in \cite{DS11}. In section \ref{sec:corfun} the correlation functions of interest and their symmetries are discussed and the Ward identities derived from the conserved charges in the double model are written down. These Ward identities are then manipulated to give the differential equations parametrising the correlation functions of primary and descendent twist fields. As the equations for the primary twist fields differ from those of the literature, section \ref{sec:analysis} presents numeric and analytic evidence that the equations derived in this paper give the correct parametrisation. Finally section \ref{sec:sum} gives our conclusions and outlook.


\section{The Model and its Fields} \label{sec:model}

In all that follows we will use the following conventions: $x\in\R$ is a space coordinate while $y\in\R$ is the usual Euclidean time. These are combined into the complex coordinates $z:=-\frac{i}{2}(x+iy)$, $\bar{z}:=\frac{i}{2}(x-iy)$, with derivatives $\partial:=\partial_{z} =i\partial_{x}+\partial_{y}$ and $\bar{\partial}:=\partial_{\bar{z}}=-i\partial_{x}+\partial_{y}$. When only one coordinate is given it is understood to be the space coordinate with time being set to $0$.

\subsection{Dirac Fermions}

The standard, two dimensional, free Dirac Fermi field of mass $m$ is an operator solution, $\Psi_R(x,y),\;\Psi_L(x,y)$, to the equations of motion
\beq\ba{ll}
 \bar{\partial}\Psi_{R}=-im\Psi_{L},& \bar{\partial}\Psi_{R}^{\dagger}=-im\Psi_{L}^{\dagger} \z
 \partial\Psi_{L}=im\Psi_{R}, & \partial\Psi_{L}^{\dagger}=im\Psi_{R}^{\dagger} \ea \label{eom}
\eeq
and the equal-time anti-commutation relations
\begin{equation}
	\{ \Psi_{R}(x_{1}),\Psi_{R}^{\dagger}(x_{2})\}=4\pi\delta(x_{1}-x_{2}) ,\quad \{\Psi_{L}(x_{1}),\Psi_{L}^{\dagger}(x_{2})\}=4\pi\delta(x_{1}-x_{2}) \label{eq:diracanticom}
\end{equation}
with all other combinations anti-commuting. This is subject to the condition there exists a vacuum state, $\rvac$, with the properties
\beq\label{defvac}
	\lim_{y\to-\infty} \Psi_{R,L}(x,y)|\vac\ket =
	\lim_{y\to-\infty} \Psi_{R,L}^\dag(x,y)|\vac\ket = 0.
\eeq
The anti-commutation relations (\ref{eq:diracanticom}) have been chosen so as to give the fields the `CFT' normalisation in terms of the $z$ variable: as $|z_1-z_2|\to 0$,
\beqa
	\bra\vac| {\cal T}\lt[\Psi_R^\dag(x_1,y_1) \Psi_R(x_2,y_2)\rt]|\vac\ket &\sim&
		\frc1{z_1-z_2},\n
	\bra\vac| {\cal T}\lt[\Psi_L^\dag(x_1,y_1) \Psi_L(x_2,y_2)\rt]|\vac\ket &\sim& \frc1{\b{z}_1-\b{z}_2}
		.\no
\eeqa
Where ${\cal T}$ is time ordering, with operators at later Euclidean time being placed to the left of those at earlier times and picking up a sign, in general, if those operators are fermionic.

These operators can be written in terms of creation and annihilation operators (Fourier modes):
\beqa
 \Psi_{R}(x,y)&=&\sqrt{m}\int{\rm d}\theta\, e^{\theta/2}\lt(D_{+}^{\dagger}(\theta)e^{y E_\theta-ix p_\theta}-i D_{-}(\theta)e^{-y E_\theta+ix p_\theta}\rt) \nonumber \\
 \Psi_{L}(x,y)&=&\sqrt{m}\int{\rm d}\theta \,e^{-\theta/2}\lt(iD_{+}^{\dagger}(\theta)e^{y E_\theta-ix p_\theta}-D_{-}(\theta)e^{-y E_\theta+ix p_\theta}\rt)\label{psi}
\eeqa
where
\beq
	E_\theta = m\cosh\theta,\quad p_\theta = m\sinh\theta
\eeq
and their Hermitian conjugates are
\beqa
 \Psi_{R}^\dag(x,y)&=&\sqrt{m}\int{\rm d}\theta\, e^{\theta/2}\lt(iD_{-}^{\dagger}(\theta)e^{y E_\theta-ix p_\theta}+D_{+}(\theta)e^{-y E_\theta+ix p_\theta}\rt) \nonumber \\
 \Psi_{L}^\dag(x,y)&=&\sqrt{m}\int{\rm d}\theta \,e^{-\theta/2}\lt(-D_{-}^{\dagger}(\theta)e^{y E_\theta-ix p_\theta}-iD_{+}(\theta)e^{-y E_\theta+ix p_\theta}\rt)\label{psidag}
\eeqa
remembering that since $y$ is Euclidean time it changes sign under Hermitian conjugation. The creation and annihilation operators satisfy the anti-commutation relations
\beq
 \{D_{+}^{\dagger}(\theta_1),D_{+}(\theta_2)\}=\delta(\theta_1-\theta_2),\quad \{D_{-}^{\dagger}(\theta_1),D_{-}(\theta_2)\}=\delta(\theta_1-\theta_2) \label{eq:diraccandaanticom}
\eeq
(with all other combinations anti-commuting), which are derived from (\ref{eq:diracanticom}), and the vacuum condition
\begin{equation}
  D_{\pm}(\theta)\rvac=0 \no
\end{equation}
from (\ref{defvac}). The multi-particle states
\begin{equation}
 |\theta_{1}\cdots\theta_{n}\rangle_{\epsilon_{1}\cdots\epsilon_{n}}:=D_{\epsilon_{1}}^{\dagger}(\theta_{1})\cdots D_{\epsilon_{n}}^{\dagger}(\theta_{n})|\vac\rangle \quad \mbox{for}\quad \theta_1>\cdots>\theta_n,\ \epsilon_{i}\in\{+,-\}. \label{basis}
\end{equation}
form a basis of the Hilbert space and have total energies $\sum_j E_{\theta_j}$ and total momenta $\sum_j p_{\theta_j}$. Dual vectors are written as ${}^{\epsilon_{1}\cdots\epsilon_{n}}\bra\theta_{1}\cdots\theta_{n}|:= |\theta_{1}\cdots\theta_{n}\rangle_{\epsilon_{1}\cdots\epsilon_{n}}^\dag$ and clearly
\begin{equation}	{}^{\epsilon_{1}'\cdots\epsilon_{n}'}\bra\theta_{1}'\cdots\theta_{n}'|\theta_{1}\cdots\theta_{n}\rangle_{\epsilon_{1}\cdots\epsilon_{n}}=\prod_{j=1}^n \delta^{\epsilon_j'}_{\epsilon_j}\delta(\theta_j-\theta_j^{\prime}).
\end{equation}
It is useful to note that the identity operator can be written as
\begin{equation} 1=\displaystyle\sum_{N=0}^{\infty}\frac{1}{N!}\displaystyle\sum_{\epsilon_{1}\cdots\epsilon_{N}}\int_{-\infty}^{\infty}d\theta_{1}\cdots\int_{-\infty}^{\infty}d\theta_{N}\,|\theta_{1}\cdots\theta_{N}\rangle_{\epsilon_{1}\cdots\epsilon_{N}}^{\phantom{\epsilon_{1}\cdots\epsilon_{N}}\epsilon_{N}\cdots\epsilon_{1}}\langle\theta_{N}\cdots\theta_{1}|. \label{eq:resofid}
\end{equation}

\subsection{Primary Twist Fields}

The internal $U(1)$ symmetry of the Dirac theory, $\Psi_{R,L}\mapsto e^{2\pi i\alpha}\Psi_{R,L}$, $\alpha\in[0,1)$, has associated primary twist fields, $\sigma_{\alpha}(x,y)$. These are local, spin-less, $U(1)$ neutral bosonic quantum fields with scaling dimension $\alpha^{2}$ \cite{MSJ78}. At this point we note that these twist fields can be identified with exponentials of the sine-Gordon field at the free fermion point via $\sigma_{\alpha}\equiv :e^{i\alpha\Phi}:$, where $\Phi$ is the appropriately normalised sine-Gordon field. The relevant bosonisation relations also give bilinears of fermions in terms of $\Phi$: $\Psi_{R}^{\dagger}\Psi_{L}\equiv :e^{i\Phi}:$ and $\Psi_{R}\Psi_{L}^{\dagger}\equiv :e^{-i\Phi}:$. Expressions for the fermions themselves require exponentials of terms which are non-local in terms of the sine-Gordon field and so the descendent fields discussed in section \ref{sec:desc-twist-fields} are also given by non-local expressions, meaning that these fields are not so natural in the sine-Gordon theory. The implications of the identification with the sine-Gordon model are beyond the scope of this paper but discussed further in \cite{BL97}.

The twist property of $\sigma_{\alpha}$ is manifested in the monodromy property of time ordered correlation functions
\begin{equation}
  C(z) := \bra\vac|{\cal T}\lt[\cdots \Psi_{R,L}(x,y)\sigma_{\alpha}(0)\cdots\rt] |\vac\ket.
\end{equation}
When the complex variable $z$ is continued along a loop $\gamma$, surrounding the origin once counter-clockwise, the function $C$ is given by
\begin{equation}
  C(e^{2\pi i}z)=e^{-2\pi i\alpha}C(z)
\end{equation}
which holds for any loop which can be contracted to the origin without intersecting the position of any other twist fields present in the correlation function. When $\Psi_{R,L}$ is replaced with $\Psi^{\dagger}_{R,L}$ a similar property holds but with $e^{-2\pi i\alpha}$ replaced by $e^{2\pi i\alpha}$. This twist property can also be expressed via the exchange relations
\begin{subequations} \label{eq:equaltimebraiding}
\beqa
 \Psi_{R,L}(x)\sigma_{\alpha}(0)&=&\left\{ \begin{array}{ll}
                                         \sigma_{\alpha}(0)\Psi_{R,L}(x) & (x<0)  \z
					 e^{2\pi i\alpha}\sigma_{\alpha}(0)\Psi_{R,L}(x) & (x>0)
                                        \end{array} \right.  \\
 \Psi_{R,L}^{\dagger}(x)\sigma_{\alpha}(0)&=&\left\{ \begin{array}{ll}
                                         \sigma_{\alpha}(0)\Psi_{R,L}^{\dagger}(x) & (x<0)  \z
					 e^{-2\pi i\alpha}\sigma_{\alpha}(0)\Psi_{R,L}^{\dagger}(x) & (x>0).
                                        \end{array} \right.
\eeqa
\end{subequations}
The Hermitian conjugates of the primary twist fields is given by
\begin{equation}\label{herm}
  \sigma_{\alpha}^{\dagger}=\sigma_{-\alpha}.
\end{equation}

From (\ref{eq:equaltimebraiding}) the form factors of the primary twist fields can be calculated \cite{K78,MSS81,EL93}. The fields are $U(1)$ neutral and so only have non-vanishing form factors with $U(1)$ neutral states:
\begin{multline} 
 \langle \vac|\sigma_{\alpha}(0)|\theta_{1}\theta_{2}\cdots\theta_{2n}\rangle_{+\cdots+-\cdots-}  =  c_{\alpha}m^{\alpha^{2}}(-1)^{n(n-1)/2}\left(\frac{\sin(\pi\alpha)}{\pi i}\right)^{n} \\ \times\ \left(\prod^{n}_{i=1}(u_{i})^{\frac{1}{2}+\alpha}(u_{i+n})^{\frac{1}{2}-\alpha}\right)  
   \frac{\prod_{i<j\leq n}(u_{i}-u_{j})\prod_{n+1\leq i<j}(u_{i}-u_{j})}{\prod_{r=1}^{n}\prod_{s=n+1}^{2n}(u_{r}+u_{s})}
 	 \label{eq:sigmaaadef}
\end{multline}
where there are $n$ $-$'s and $n$ $+$'s in the state, and $u_{i}:=\exp(\theta_{i})$. In particular, the two-particle form factor is
\beq\label{2pff}
	\bra\vac|\sigma_\alpha(0)|\theta_1\theta_2\ket_{+-} = c_\alpha m^{\alpha^2}
	\frc{\sin(\pi\alpha)}{2\pi i} \frc{e^{\alpha(\theta_1-\theta_2)}}{\cosh\frc{\theta_1-\theta_2}2}.
\eeq
These form factors may differ from those written in other publications in the choice of sign of $\alpha$ and the two particle form factor. The constant $c_{\alpha}$ does not influence the calculations of this paper and so we direct the interested reader to \cite{BT92}, \cite{LZ96}, \cite{D03} and \cite{DS11} for further discussion and evaluation of this constant.

Other matrix elements can be obtained by analytic continuation in the rapidities, as is standard in the context of 1+1-dimensional QFT \cite{S92}. Note that the form factors (\ref{eq:sigmaaadef}) and the Hermiticity relation (\ref{herm}) are in agreement with the analytic-continuation formula
\beq
	\langle \vac|\sigma_{\alpha}(0)|(\theta_{1}+i\pi)\cdots(\theta_{2n}+i\pi)\rangle_{+\cdots+-\cdots-}
	={}^{+\cdots+-\cdots-}\bra \theta_{2n} \cdots \theta_1|\sigma_{\alpha}(0)|\vac\ket,
\eeq
where the analytic continuation is simultaneous on all rapidities.

\subsection{Descendent Twist Fields}\label{sec:desc-twist-fields}

As well as the primary twist field described in the previous subsection our model also contains descendent twist fields. The two families of descendent twist field of interest here occur in the operator product expansions of primary twist field with the Dirac fields $\Psi_{R}$ and $\Psi_{R}^{\dagger}$:
\begin{eqnarray}
  \T\left[\Psi_{R}^{\dagger}(x,y)\sigma_{\alpha}(0)\right] & \sim & (-iz)^{\alpha}\sigma_{\alpha+1,\alpha}(0) \n
  \T\left[\Psi_{R}(x,y)\sigma_{\alpha}(0)\right] & \sim & (-iz)^{-\alpha}\sigma_{\alpha-1,\alpha}(0).
\label{eq:ope}
\end{eqnarray}
These OPEs are valid for all $\alpha\in\R\backslash\Z^{*}$ (where $\Z^{*}:=\Z\backslash\{0\}$) and the same families of descendent twist fields occur in the OPEs of $\Psi_{L}$ and $\Psi_{L}^{\dagger}$ with primary twist fields:
\begin{eqnarray}
  \T\left[\Psi_{L}^{\dagger}(x,y)\sigma_{\alpha}(0)\right] & \sim & -(i\bar{z})^{-\alpha}\sigma_{\alpha,\alpha-1}(0) \n
  \T\left[\Psi_{L}(x,y)\sigma_{\alpha}(0)\right] & \sim & -(i\bar{z})^{\alpha}\sigma_{\alpha,\alpha+1}(0).
\end{eqnarray}
The descendent twist fields $\sigma_{\alpha\pm1,\alpha}$ have dimensions $\alpha^{2}\pm\alpha+1/2$, spins $\pm\alpha+1/2$ and charges $\mp1$, as discussed in \cite{DS11}. The one particle form factors of these fields can be evaluated from the OPEs:
\begin{eqnarray}
 \langle\vac|\sigma_{\alpha+1,\alpha}(0)|\theta\rangle_{+} & = & c_{\alpha}\frc{e^{-i\pi\alpha/2}}{\Gamma(1+\alpha)}m^{\alpha^{2}+\alpha+1/2}e^{(\alpha+1/2)\theta} \label{eq:1pff(a+1,a)} \\
 \langle\vac|\sigma_{\alpha-1,\alpha}(0)|\theta\rangle_{-}  & = & -ic_{\alpha}\frac{e^{i\pi\alpha/2}}{\Gamma(1-\alpha)}m^{\alpha^{2}-\alpha+1/2}e^{(-\alpha+1/2)\theta}.  \label{eq:1pff(a-1,a)}
\end{eqnarray}
All higher particle form factors can be calculated via Wick's theorem.


\section{The Double Model} \label{sec:DoubleMod}

As was the case in \cite{DS11} inspiration is taken from \cite{FZ03} where a model containing two non-interacting copies of Ising field theory is studied. The conserved charges in this double model allows differential equations for the Ising spin-spin correlation function to be written down.

Here we consider a double model consisting of two non-interacting copies of the Dirac model discussed in section \ref{sec:model} with $\Psi$ and $\Phi$ denoting the fermion fields and $D_{\pm}^{(\dagger)}$ and $E_{\pm}^{(\dagger)}$ the respective creation and annihilation operators of the two copies. As these two copies do not interact their fermionic fields anti-commute. The copy which other fields belong to will be identified by a superscript; so twist fields from the two copies are denoted $\sigma_{\alpha}^{\Psi}$ and $\sigma_{\alpha}^{\Phi}$.

\subsection{Conserved charges}

Energy and momentum are conserved charges of the single-copy model. They are associated to the dynamical invariance under space and time translation: the equations of motion (\ref{eom}) possesses this invariance. In the double-copy model, the energy and momentum conserved charges are the sums of the corresponding charges of each copy. However, there are other conserved charges in the double-copy model that can be constructed out of these. As the two copies are non-interacting, the energy-momentum operators for each are still independently conserved quantities, and can be combined differently to give new conserved quantities. The two specific conserved charges that are of interest to us are the differences of those of the single copies. We define $P$ and $\bar{P}$ via the following action (these conserved charges are chosen to be anti-Hermitian):
\begin{align}
 [P,\mathcal{O}^{\Psi}\mathcal{O}^{\Phi}]=& i\partial\mathcal{O}^{\Psi}\mathcal{O}^{\Phi}-i\mathcal{O}^{\Psi}\partial\mathcal{O}^{\Phi} \nonumber \\
 [\bar{P},\mathcal{O}^{\Psi}\mathcal{O}^{\Phi}]=& -i\bar{\partial}\mathcal{O}^{\Psi}\mathcal{O}^{\Phi}+i\mathcal{O}^{\Psi}\bar{\partial}\mathcal{O}^{\Phi}. \label{eq:actionPPbar}
\end{align}
This holds for any local fields $\mathcal{O}^{\Psi}$ and $\mathcal{O}^{\Phi}$ interacting non-trivially with fields in copy $\Psi$ and $\Phi$ respectively. In particular, the actions on the creation and annihilation operators are
\beqa
	&& [P,D_\pm(\theta)] = -ime^\theta D_\pm(\theta),\quad [P,E_\pm(\theta)] = ime^\theta E_\pm(\theta),\n
	&& [\b{P},D_\pm(\theta)] = ime^{-\theta} D_\pm(\theta),\quad [\b{P},E_\pm(\theta)] = -ime^{-\theta} E_\pm(\theta) \no
\eeqa
from which it is simple to derive an explicit expression for $P$ and $\b{P}$ through bilinears in the creation and annihilation operators.

In the double-copy model, there is another conserved charge $Z$, related to the $O(2)$ rotation symmetry amongst the copies. Written in terms of the Fermi fields $\Psi$ and $\Phi$, the charge $Z$ is (again, chosen to be anti-Hermitian)
\begin{equation}
 Z=\frac{1}{4\pi}\int{\rm d}x(\Psi_{R}\Phi_{R}^{\dagger}+\Psi_{L}\Phi_{L}^{\dagger}+\Psi_{R}^{\dagger}\Phi_{R}+\Psi_{L}^{\dagger}\Phi_{L}). \label{eq:defZ}
\end{equation}
The action of this charge on the creation and annihilation operators is
\begin{eqnarray}
 [Z,D_{\pm}^{\dagger}(\theta)]=-E_{\pm}^{\dagger}(\theta) & [Z,E_{\pm}^{\dagger}(\theta)]=D_{\pm}^{\dagger}(\theta) \label{eq:actionZ}
\end{eqnarray}
and similarly for $D_{\pm}(\theta)$ and $E_{\pm}(\theta)$.

Higher order conserved charges can of course be obtained by commutations of $P$, $\b{P}$ and $Z$ but as was the case in \cite{DS11} the Ward identities coming from these three charges prove sufficient to derive the equations of interest. Below, we will derive the integrable differential equations from the Ward identities associated to $Z$, and use the simpler space-time dependence of correlation functions coming from translation, parity and rotation symmetries to achieve our result.

As was observed in \cite{Dth} (in the generalised situation of a theory on the Poincar\'e disk), it is the relation
\begin{align}\label{eqmass}
 [P,[\bar{P},Z]]=[\bar{P},[P,Z]]=4m^{2}Z
\end{align}
which provides the dependence on the mass in the differential equations. Relation (\ref{eqmass}) corresponds essentially to the equations of motion of the theory, as expressed using the charge $Z$. It can be derived from (\ref{eq:actionPPbar}) and (\ref{eq:defZ}) using the equations of motion (\ref{eom}). Relation (\ref{eqmass}) is the only one where the massive theory is used: besides it, we only need the action of $Z$ on products of twist fields and their first derivatives in order to find differential equations parametrising the correlation function (\ref{eq:defSig}).

\subsection{Equations for the action of $Z$}\label{sec:eqZ}

The action of the conserved charge $Z$ on products of local fields $\mathcal{O}^{\Psi}\mathcal{O}^{\Phi}$ gives linear combinations of similar products of local fields (with unchanged locality index) whenever the combined locality index of the product $\mathcal{O}^{\Psi}\mathcal{O}^{\Phi}$ is one. Otherwise, the result of the action is non-local, and the resulting Ward identities are not useful. The equations that are needed here are derived in \cite{DS11} so we simply report the necessary results.

In the rest of this section we neglect to write the location of the fields, to make the equations easier to read, it being understood that all fields are evaluated at the same point, $(0,0)$ for instance. The equations representing the action of the charge $Z$ on twist fields involving no derivatives are:
\begin{eqnarray*}
 [Z,\sigma_{\alpha}^{\Psi}\sigma_{\alpha}^{\Phi}] &= & 0  \label{eq:[Z,aa]} \\
{} [Z,\sigma_{\alpha}^{\Psi}\sigma_{\alpha-1}^{\Phi}] &= & i(\sigma_{\alpha-1,\alpha}^{\Psi}\sigma_{\alpha,\alpha-1}^{\Phi}-\sigma_{\alpha,\alpha-1}^{\Psi}\sigma_{\alpha-1,\alpha}^{\Phi}) \label{eq:[Z,aa-1]} \\
{} [Z,\sigma_{\alpha}^{\Psi}\sigma_{\alpha+1}^{\Phi}] &= & i(\sigma_{\alpha+1,\alpha}^{\Psi}\sigma_{\alpha,\alpha+1}^{\Phi}-\sigma_{\alpha,\alpha+1}^{\Psi}\sigma_{\alpha+1,\alpha}^{\Phi}) \label{eq:[Z,aa+1]} \\
{} [Z,\sigma_{\alpha+1,\alpha}^{\Psi}\sigma_{\alpha,\alpha+1}^{\Phi}] &= & i( \sigma_{\alpha}^{\Psi}\sigma_{\alpha+1}^{\Phi}-\sigma_{\alpha+1}^{\Psi}\sigma_{\alpha}^{\Phi}). \label{eq:[Z,mix]} 
\end{eqnarray*}

Equations involving one derivative are as follows:
\begin{eqnarray*}
{} [Z,(\partial\sigma_{\alpha}^{\Psi})\sigma_{\alpha+1}^{\Phi}] &= & i((\partial\sigma_{\alpha,\alpha+1}^{\Psi})\sigma_{\alpha+1,\alpha}^{\Phi}-\sigma_{\alpha+1,\alpha}^{\Psi}(\partial\sigma_{\alpha,\alpha+1}^{\Phi}))  \\
{} [Z,(\bar{\partial}\sigma_{\alpha}^{\Psi})\sigma_{\alpha+1}^{\Phi}] &= & i(\sigma_{\alpha,\alpha+1}^{\Psi}(\bar{\partial}\sigma_{\alpha+1,\alpha}^{\Phi})-(\bar{\partial}\sigma_{\alpha+1,\alpha}^{\Psi})\sigma_{\alpha,\alpha+1}^{\Phi})  \\
{} [Z,(\partial\sigma_{\alpha}^{\Psi})\sigma_{\alpha-1}^{\Phi}] &= & i((\partial\sigma_{\alpha,\alpha-1}^{\Psi})\sigma_{\alpha-1,\alpha}^{\Phi}-\sigma_{\alpha-1,\alpha}^{\Psi}(\partial\sigma_{\alpha,\alpha-1}^{\Phi}))  \\
{} [Z,(\bar{\partial}\sigma_{\alpha}^{\Psi})\sigma_{\alpha-1}^{\Phi}] &= & i(\sigma_{\alpha,\alpha-1}^{\Psi}(\bar{\partial}\sigma_{\alpha-1,\alpha}^{\Phi})-(\bar{\partial}\sigma_{\alpha-1,\alpha}^{\Psi})\sigma_{\alpha,\alpha-1}^{\Phi})  \\
{} [Z,(\partial\sigma_{\alpha,\alpha+1}^{\Psi})\sigma_{\alpha+1,\alpha}^{\Phi}] &= & i(\sigma_{\alpha+1}^{\Psi}(\partial\sigma_{\alpha}^{\Phi})-(\partial\sigma_{\alpha}^{\Psi})\sigma_{\alpha+1}^{\Phi})  \\
{} [Z,(\partial\sigma_{\alpha+1,\alpha}^{\Psi})\sigma_{\alpha,\alpha+1}^{\Phi}]&= & i(\sigma_{\alpha}^{\Psi}(\partial\sigma_{\alpha+1}^{\Phi})-(\partial\sigma_{\alpha+1}^{\Psi})\sigma_{\alpha}^{\Phi})  \\
{} [Z,(\bar{\partial}\sigma_{\alpha,\alpha+1}^{\Psi})\sigma_{\alpha+1,\alpha}^{\Phi}]&= & i((\bar{\partial}\sigma_{\alpha+1}^{\Psi})\sigma_{\alpha}^{\Phi}-\sigma_{\alpha}^{\Psi}(\bar{\partial}\sigma_{\alpha+1}^{\Phi})) \\
{} [Z,(\bar{\partial}\sigma_{\alpha+1,\alpha}^{\Psi})\sigma_{\alpha,\alpha+1}^{\Phi}]&= & i((\bar{\partial}\sigma_{\alpha}^{\Psi})\sigma_{\alpha+1}^{\Phi}-\sigma_{\alpha+1}^{\Psi}(\bar{\partial}\sigma_{\alpha}^{\Phi})).
\end{eqnarray*}

The only relations involving second derivatives of twist fields that we need can be obtained by using those above along with the equation of motion of the charge $Z$ (\ref{eqmass}). Indeed, we only need action of $Z$ on double-derivative fields of the form $[P,[\bar{P},\mathcal{O}^{\Psi}\mathcal{O}^{\Phi}]]$, which we can evaluate using
\begin{align*}
	\lefteqn{[Z,[P,[\bar{P},\mathcal{O}^{\Psi}\mathcal{O}^{\Phi}]]]} & \\ & =
	[[P,[\bar{P},Z]],\mathcal{O}^{\Psi}\mathcal{O}^{\Phi}]+
	[\bar{P},[Z,[P,\mathcal{O}^{\Psi}\mathcal{O}^{\Phi}]]]+
	[P,[Z,[\bar{P},\mathcal{O}^{\Psi}\mathcal{O}^{\Phi}]]]-
	[P,[\bar{P},[Z,\mathcal{O}^{\Psi}\mathcal{O}^{\Phi}]]] \\ &=
	4m^2 [Z,\mathcal{O}^{\Psi}\mathcal{O}^{\Phi}]+
	[\bar{P},[Z,[P,\mathcal{O}^{\Psi}\mathcal{O}^{\Phi}]]]+
	[P,[Z,[\bar{P},\mathcal{O}^{\Psi}\mathcal{O}^{\Phi}]]]-
	[P,[\bar{P},[Z,\mathcal{O}^{\Psi}\mathcal{O}^{\Phi}]]].
\end{align*}
This is how the mass dependence will appear in the equations for correlation functions.


\section{Correlation Functions} \label{sec:corfun}

The aim of this section is to generalise the results of \cite{DS11} and obtain a pair of partial differential equations parametrising the correlation function 
\begin{equation}
  \bra\sigma_{\alpha}(x,y)\sigma_{\beta}(0,0)\ket
\end{equation}
using the Ward identities derived from the action of $Z$. 

Before describing the properties of the correlation functions we first introduce the notation
\begin{equation}
  \label{eq:Fabcd}
  F_{\alpha,\beta}^{\gamma,\delta}(x,y)=\bra\sigma_{\alpha,\beta}(x,y)\sigma_{\gamma,\delta}(0,0)\ket
\end{equation}
where it is understood that when the two indices are equal the twist field is a primary twist field, i.e.
\begin{equation*}
  \sigma_{\alpha,\alpha}:=\sigma_{\alpha}.
\end{equation*}

\subsection{Symmetries of Correlation Functions} \label{sec:symcorfunc}

Before writing down our Ward identities we will first discuss some symmetries of the correlation functions involved. Firstly since our system has translation symmetry it is always possible to place one of the fields at the origin, as in (\ref{eq:Fabcd}).

Next we note that 
\begin{equation}
  F_{\alpha+1,\alpha+1}^{\beta+1,\beta+1}(x,y)=F_{\alpha,\alpha}^{\beta,\beta}(x,y)
\end{equation}
which follows directly from the definition of the primary twist fields and can be seen explicitly by inserting the identity (\ref{eq:resofid}) between the fields on each side and using the form factors (\ref{eq:sigmaaadef}). We also note that as the fields involved in $F_{\alpha,\alpha}^{\beta,\beta}$ are spin-less we do not expect to this function to depend on the angle $\theta$, where $z=re^{i\theta}$ and $\bar{z}=re^{-i\theta}$, so we may also write
\begin{equation}
  F_{\alpha,\alpha}^{\beta,\beta}(x,y)=F_{\alpha,\alpha}^{\beta,\beta}(r).
\end{equation}

Along with translation symmetry we also assume parity symmetry, discussed in more detail in \cite{DS11}. Combining this with the fact that the primary twist fields have bosonic statistics, we see that $F_{\alpha,\alpha}^{\beta,\beta}(x,y)= F_{\beta,\beta}^{\alpha,\alpha}(x,y)$. In order to achieve a similar relation for the descendent twist fields we need to return to the braiding relations (\ref{eq:equaltimebraiding}) and recall that by definition 
\begin{align}
 \sigma_{\alpha+1,\alpha}(w)=\lim_{z\rightarrow w}(-i(z-w))^{-\alpha}\Psi_{R}^{\dagger}(z)\sigma_{\alpha,\alpha}(w) \nonumber \\
 \sigma_{\alpha,\alpha+1}(w)=\lim_{z\rightarrow w}(i(\bar{z}-\bar{w}))^{-\alpha}\Psi_{L}(z)\sigma_{\alpha,\alpha}(w) 
\end{align}
where the factors $(-i(z-w))^{-\alpha}$ and $(i(\bar{z}-\bar{w}))^{-\alpha}$ are taken on the principal branch and so are continuous exactly where $\Psi_{R}^{\dagger}(z)\sigma_{\alpha,\alpha}(w)$ and $\Psi_{L}(z)\sigma_{\alpha,\alpha}(w)$ are continuous. Using these definitions and the braiding relations we obtain
\begin{align}
 \langle\sigma_{\beta,\beta+1}(w_{2})\sigma_{\alpha+1,\alpha}(w_{1})\rangle=\left\{ \begin{array}{cc}
                                        			 -e^{2\pi i\beta} & {\rm if }~ x(w_{1})>x(w_{2})  \\
					 			 -e^{2\pi i\alpha} & {\rm if }~ x(w_{1})<x(w_{2})
                                        		       \end{array} \right\} \langle\sigma_{\alpha+1,\alpha}(w_{1})\sigma_{\beta,\beta+1}(w_{2})\rangle.
\end{align}
We can now write
\begin{align}
 \langle\sigma_{\alpha+1,\alpha}(w_{1})\sigma_{\beta,\beta+1}(w_{2})\rangle=(i(w_{1}-w_{2}))^{\beta-\alpha}f_{\alpha+1,\alpha}^{\beta,\beta+1}(|w_{1}-w_{2}|) \label{eq:twistfuncdep1}
\end{align}
which reproduces the correct rotation properties. The choice of the factor $(i(w_{1}-w_{2}))^{\beta-\alpha}$ is in some sense arbitrary, but defined on the principal branch, it is convenient as we can also write
\begin{equation}
 \langle\sigma_{\beta,\beta+1}(w_{2})\sigma_{\alpha+1,\alpha}(w_{1})\rangle=-e^{2\pi i\zeta}(i(w_{1}-w_{2}))^{\beta-\alpha}f_{\alpha+1,\alpha}^{\beta,\beta+1}(|w_{1}-w_{2}|) \label{eq:twistfuncdep2}
\end{equation}
where the same function $f_{\alpha+1,\alpha}^{\beta,\beta+1}(|w_{1}-w_{2}|)$ is involved and $\zeta=\alpha$ if $x(w_{1})<x(w_{2})$ and $\zeta=\beta$ if $x(w_{1})>x(w_{2})$. This choice will not be a problem as we will always consider products of the form $F_{\alpha,\alpha+1}^{\beta+1,\beta}F_{\alpha+1,\alpha}^{\beta,\beta+1}$ and so the same constant will occur in every term.

With all this in mind we can now write
\begin{eqnarray}
  F_{\alpha,\alpha+1}^{\beta+1,\beta}(x,y)=pe^{i\theta(\alpha-\beta)}f_{1}(r)  \n
  F_{\alpha+1,\alpha}^{\beta,\beta+1}(x,y)=qe^{i\theta(\beta-\alpha)}f_{2}(r)  \label{eq:funcdesctwist}
\end{eqnarray}
where $p$ and $q$ are phases, as in (\ref{eq:twistfuncdep2}), coming from (\ref{eq:equaltimebraiding}), and the functions $f_{1}$ and $f_{2}$ will be discussed in the next subsection.

\subsection{Ward Identities} \label{sec:wardidents}

We are now in a position to write down the Ward identities associated with the actions of $P$, $\bar{P}$ and $Z$ and simplify the resulting equations.

The Ward identities of interest here are:
\begin{subequations}\label{eq:wards1}
 \begin{align}
  \langle[Z,\sigma_{\alpha,\alpha}^{\Psi}(z)\sigma_{\alpha+1,\alpha+1}^{\Phi}(z)\sigma_{\beta+1,\beta}^{\Psi}(0)\sigma_{\beta,\beta+1}^{\Phi}(0)]\rangle=0 \\
  \langle[Z,[P,\sigma_{\alpha,\alpha}^{\Psi}(z)\sigma_{\alpha+1,\alpha+1}^{\Phi}(z)]\sigma_{\beta+1,\beta}^{\Psi}(0)\sigma_{\beta,\beta+1}^{\Phi}(0)]\rangle=0 \\
  \langle[Z,[\bar{P},\sigma_{\alpha,\alpha}^{\Psi}(z)\sigma_{\alpha+1,\alpha+1}^{\Phi}(z)]\sigma_{\beta+1,\beta}^{\Psi}(0)\sigma_{\beta,\beta+1}^{\Phi}(0)]\rangle=0 \\
  \langle[Z,[P,[\bar{P},\sigma_{\alpha,\alpha}^{\Psi}(z)\sigma_{\alpha+1,\alpha+1}^{\Phi}(z)]]\sigma_{\beta+1,\beta}^{\Psi}(0)\sigma_{\beta,\beta+1}^{\Phi}(0)]\rangle=0 \\
  \langle[Z,[P,\sigma_{\alpha,\alpha}^{\Psi}(z)\sigma_{\alpha+1,\alpha+1}^{\Phi}(z)][\bar{P},\sigma_{\beta+1,\beta}^{\Psi}(0)\sigma_{\beta,\beta+1}^{\Phi}(0)]]\rangle=0.
 \end{align}
\end{subequations}
Maintaining the full $(z,\bar{z})$ dependence for the moment these identities become, respectively,
\begin{subequations}
 \begin{align}
  F_{\alpha,\alpha+1}^{\beta+1,\beta}(z)F_{\alpha+1,\alpha}^{\beta,\beta+1}(z)-F_{\alpha,\alpha}^{\beta+1,\beta+1}(z)F_{\alpha+1,\alpha+1}^{\beta,\beta}(z)+F_{\alpha,\alpha}^{\beta,\beta}(z)F_{\alpha+1,\alpha+1}^{\beta+1,\beta+1}(z)=0 \label{eq:ward1B}
 \end{align}
 \begin{multline}
 \partial F_{\alpha,\alpha+1}^{\beta+1,\beta}(z)F_{\alpha+1,\alpha}^{\beta,\beta+1}(z)-F_{\alpha,\alpha+1}^{\beta+1,\beta}(z)\partial F_{\alpha+1,\alpha}^{\beta,\beta+1}(z)+ \\ \partial F_{\alpha,\alpha}^{\beta+1,\beta+1}(z)F_{\alpha+1,\alpha+1}^{\beta,\beta}(z) -F_{\alpha,\alpha}^{\beta+1,\beta+1}(z)\partial F_{\alpha+1,\alpha+1}^{\beta,\beta}(z)=0 \label{eq:ward2B}
  \end{multline}
  \begin{multline}
 \bar{\partial}F_{\alpha,\alpha+1}^{\beta+1,\beta}(z)F_{\alpha+1,\alpha}^{\beta,\beta+1}(z)-F_{\alpha,\alpha+1}^{\beta+1,\beta}(z)\bar{\partial}F_{\alpha+1,\alpha}^{\beta,\beta+1}(z) \\ -\bar{\partial}F_{\alpha,\alpha}^{\beta+1,\beta+1}(z)F_{\alpha+1,\alpha+1}^{\beta,\beta}(z) +F_{\alpha,\alpha}^{\beta+1,\beta+1}(z)\bar{\partial}F_{\alpha+1,\alpha+1}^{\beta,\beta}(z) =0 \label{eq:ward3B}
  \end{multline}
 \begin{multline}
  \partial\bar{\partial}F_{\alpha,\alpha+1}^{\beta+1,\beta}(z)F_{\alpha+1,\alpha}^{\beta,\beta+1}(z) -\partial F_{\alpha,\alpha+1}^{\beta+1,\beta}(z)\bar{\partial}F_{\alpha+1,\alpha}^{\beta,\beta+1}(z) -\bar{\partial}F_{\alpha,\alpha+1}^{\beta+1,\beta}(z)\partial F_{\alpha+1,\alpha}^{\beta,\beta+1}(z) \\ +F_{\alpha,\alpha+1}^{\beta+1,\beta}(z)\partial\bar{\partial}F_{\alpha+1,\alpha}^{\beta,\beta+1}(z)-4m^{2}F_{\alpha,\alpha+1}^{\beta+1,\beta}(z)F_{\alpha+1,\alpha}^{\beta,\beta+1}(z) +\partial\bar{\partial}F_{\alpha,\alpha}^{\beta+1,\beta+1}(z)F_{\alpha+1,\alpha+1}^{\beta,\beta}(z) \\ -\partial F_{\alpha,\alpha}^{\beta+1,\beta+1}(z)\bar{\partial}F_{\alpha+1,\alpha+1}^{\beta,\beta}(z) -\bar{\partial}F_{\alpha,\alpha}^{\beta+1,\beta+1}(z)\partial F_{\alpha+1,\alpha+1}^{\beta,\beta}(z) +F_{\alpha,\alpha}^{\beta+1,\beta+1}(z)\partial\bar{\partial}F_{\alpha+1,\alpha+1}^{\beta,\beta}(z) \\ -\partial\bar{\partial}F_{\alpha,\alpha}^{\beta,\beta}(z)F_{\alpha+1,\alpha+1}^{\beta+1,\beta+1}(z) +\partial F_{\alpha,\alpha}^{\beta,\beta}(z)\bar{\partial}F_{\alpha+1,\alpha+1}^{\beta+1,\beta+1}(z) +\bar{\partial}F_{\alpha,\alpha}^{\beta,\beta}(z)\partial F_{\alpha+1,\alpha+1}^{\beta+1,\beta+1}(z) \\ -F_{\alpha,\alpha}^{\beta,\beta}(z)\partial\bar{\partial}F_{\alpha+1,\alpha+1}^{\beta+1,\beta+1}(z)=0 \label{eq:ward4B}
 \end{multline}
 \begin{multline}
  \partial\bar{\partial}F_{\alpha,\alpha+1}^{\beta+1,\beta}(z)F_{\alpha+1,\alpha}^{\beta,\beta+1}(z) -\partial F_{\alpha,\alpha+1}^{\beta+1,\beta}(z)\bar{\partial}F_{\alpha+1,\alpha}^{\beta,\beta+1}(z) -\bar{\partial}F_{\alpha,\alpha+1}^{\beta+1,\beta}(z)\partial F_{\alpha+1,\alpha}^{\beta,\beta+1}(z) \\ +F_{\alpha,\alpha+1}^{\beta+1,\beta}(z)\partial\bar{\partial}F_{\alpha+1,\alpha}^{\beta,\beta+1}(z) +\partial\bar{\partial}F_{\alpha,\alpha}^{\beta,\beta}(z)F_{\alpha+1,\alpha+1}^{\beta+1,\beta+1}(z) -\partial F_{\alpha,\alpha}^{\beta,\beta}(z)\bar{\partial}F_{\alpha+1,\alpha+1}^{\beta+1,\beta+1}(z) \\ - \bar{\partial}F_{\alpha,\alpha}^{\beta,\beta}(z)\partial F_{\alpha+1,\alpha+1}^{\beta+1,\beta+1}(z) +F_{\alpha,\alpha}^{\beta,\beta}(z)\partial\bar{\partial}F_{\alpha+1,\alpha+1}^{\beta+1,\beta+1}(z) +\partial\bar{\partial}F_{\alpha,\alpha}^{\beta+1,\beta+1}(z)F_{\alpha+1,\alpha+1}^{\beta,\beta}(z) \\ -\partial F_{\alpha,\alpha}^{\beta+1,\beta+1}(z)\bar{\partial}F_{\alpha+1,\alpha+1}^{\beta,\beta}(z) -\bar{\partial}F_{\alpha,\alpha}^{\beta+1,\beta+1}(z)\partial F_{\alpha+1,\alpha+1}^{\beta,\beta}(z) \\ +F_{\alpha,\alpha}^{\beta+1,\beta+1}(z)\partial\bar{\partial}F_{\alpha+1,\alpha+1}^{\beta,\beta}(z)=0. \label{eq:ward5B}
 \end{multline}
\end{subequations}
Now switching to polar coordinates and inserting the expressions (\ref{eq:funcdesctwist}) into (\ref{eq:ward2B}) and (\ref{eq:ward3B}) we find that $f_{1}=f_{2}$ up to an irrelevant constant. Thus we let
\begin{equation}
  \label{eq:deff}
  f(r):=f_{1}(r)=f_{2}(r).
\end{equation}

We now depart from the method of \cite{DS11}; in the present case we have no information about the relation between $F_{\alpha,\alpha}^{\beta+1,\beta+1}$ and $F_{\alpha+1,\alpha+1}^{\beta,\beta}$ so we will use (\ref{eq:ward1B},\ref{eq:ward2B},\ref{eq:ward3B}) to eliminate these functions from (\ref{eq:ward4B}) and (\ref{eq:ward5B}). In doing this we will also simplify our notation, using $f$ as in (\ref{eq:deff}) and (\ref{eq:funcdesctwist}), and letting
\begin{equation}
  F_{\alpha,\alpha}^{\beta,\beta}(r)=F(r).
\end{equation}
It is also simpler to switch to $r$ and $\theta$ coordinates, as the $\theta$ dependence will drop out, and introduce the operator
\begin{equation}
  \D=\p^{2}_{r}+\frac{1}{r}\p_{r}.
\end{equation}

Putting all this together we see that (\ref{eq:ward4B}) becomes
\begin{subequations}
\begin{multline}\label{eq:ab1}
 -pqf\mathcal{D}f+pq\frac{(\alpha-\beta)^{2}}{r^{2}}f^{2}+pqm^{2}f^{2}+(\partial_{r}F)^{2} \\
 -\frac{1}{F^{2}-pqf^{2}}\left(F^{2}(\partial_{r}F)^{2}-2pqFf\partial_{r}F\partial_{r}f+(pq)^{2}f^{2}(\partial_{r}f)^{2}-(pq)^{2}(\alpha-\beta)^{2}f^{4}/r^{2}\right)=0.
\end{multline}
and similarly (\ref{eq:ward5B}) becomes
\begin{multline}\label{eq:ab2}
 -pqf\mathcal{D}f+pq\frac{(\alpha-\beta)^{2}}{r^{2}}f^{2} +F\mathcal{D}F \\  -\frac{1}{F^{2}-pqf^{2}}\left(F^{2}(\partial_{r}F)^{2}-2pqFf\partial_{r}F\partial_{r}f+(pq)^{2}f^{2}(\partial_{r}f)^{2}-(pq)^{2}(\alpha-\beta)^{2}f^{4}/r^{2}\right)=0.
\end{multline}
\end{subequations}

Now as in the case for $\alpha=\beta$ we can make progress by setting
\begin{eqnarray}
 F+\sqrt{pq}f=e^{\chi}\cosh(\varphi) & \qquad & F-\sqrt{pq}f=e^{\chi}\sinh(\varphi)
\end{eqnarray}
which reduces (\ref{eq:ab1}) and (\ref{eq:ab2}), respectively, to:
\begin{subequations}
 \begin{multline}\label{eq:ab5}
 (\mathcal{D}\chi-\mathcal{D}\varphi)(\sinh(2\varphi)-\cosh(2\varphi)) \\
 +(\partial_{r}\varphi)^{2}(2\sinh(2\varphi)-\frac{(\cosh^{2}\varphi+\sinh^{2}\varphi)^{2}}{\cosh\varphi\sinh\varphi}) \\
 +m^{2}(\cosh(2\varphi)-\sinh(2\varphi))\\ 
 +\frac{(\alpha-\beta)^{2}}{r^{2}}(\cosh(2\varphi)-\sinh(2\varphi))(1+\frac{1}{4}(\coth\varphi-1)(1-\tanh\varphi))=0
\end{multline}
\begin{multline}\label{eq:ab6}
 \sinh(2\varphi)\D\chi+\cosh(2\varphi)\D\varphi \\
 +(\pr\varphi)^{2}(\sinh(2\varphi)-\cosh(2\varphi)\coth(2\varphi)) \\
 +\frac{(\alpha-\beta)^{2}}{4r^{2}}(\cosh(2\varphi)-\sinh(2\varphi))(\coth(2\varphi)+1)=0.
\end{multline}
\end{subequations}
Eliminating $\chi$ from these equations leaves
\begin{multline}\label{eq:ab8}
 \D\varphi-(\pr\varphi)^{2}(1+\coth(2\varphi))-m^{2}\sinh(2\varphi)(\sinh(2\varphi)-\cosh(2\varphi)) \\
 +\frac{(\alpha-\beta)^{2}}{4r^{2}}(1+\coth(2\varphi))=0
\end{multline}
which we attempt to transform into (\ref{eq:result2}) meaning we must eliminate the $\pr\varphi$ term. This is accomplished by noting that
\begin{equation}\label{eq:diffpsi}
 \frac{h''}{h'}=-1-\coth(2\varphi)
\end{equation}
is solved by
\begin{equation}\label{eq:psih}
 2\psi=h(\varphi)=\ln\left(\frac{\sqrt{1-e^{4\varphi}}-1}{\sqrt{1-e^{4\varphi}}+1}\right)
\end{equation}
and it is then a straight forward exercise to show that
\begin{equation}\label{eq:myres2}
 \D\psi=\frac{m^{2}}{2}\sinh(2\psi)+\frac{(\alpha-\beta)^{2}}{r^{2}}\tanh(\psi)\left(1-\tanh^{2}(\psi)\right).
\end{equation}

To obtain (\ref{eq:result1}) we note that by setting
\begin{equation}
  e^{\Sigma}=2F(r)=e^{\chi+\varphi}
\end{equation}
we find
\begin{eqnarray}
 \D\Sigma & = & \D\chi+\D\varphi \n
	  & = & \frac{m^{2}}{2}\left(1-\cosh(2\psi)\right). \label{eq:myres1}
\end{eqnarray}

As noted in the introduction (\ref{eq:myres2}) disagrees with the equations in \cite{BL97} by a factor of $4$ in the $(\alpha-\beta)^{2}$ term, which is not just a choice of coordinates since $m$ scales like $1/r$. Evidence that (\ref{eq:myres2}) gives the correct parametrisation is presented in the next section.

It is also worth noting that by setting
\begin{equation}
  e^{\Sigma'}=2\sqrt{pq}f(r)=e^{\chi-\varphi}
\end{equation}
we find that $f$ is also parametrised by the same function, $\psi$, via the equation
\begin{equation}
  \D\Sigma'=\frac{(\pr\psi)^{2}}{\sinh^{2}\psi}-m^{2}-\frac{(\alpha-\beta)^{2}}{r^{2}\cosh^{2}\psi}
\end{equation}
which, as stated in the introduction, is a new result that appears naturally when applying this method.


\section{Analysis of Differential Equations} \label{sec:analysis}

In order to show that the results of the previous section give the correct parametrisation of the correlation functions we examine the form factor expansion of the correlator
\begin{equation}
  \label{eq:corrsig}
  \bra\sigma_{\alpha}(x,y)\sigma_{\beta}(0,0)\ket
\end{equation}
at large distances and look at how well this solves the differential equations (\ref{eq:myres2}). It turns out to be sufficient to only consider terms up to two particles so we use the expression for the two particle form factor given in (\ref{2pff}).

To begin we insert the resolution of the identity (\ref{eq:resofid}) in between the fields of (\ref{eq:corrsig}) and then expand the two particle term:
\begin{eqnarray}
 \bra\sigma_{\alpha}(x,y)\sigma_{\beta}(0,0)\ket & = & \bra\vac|\sigma_{\alpha}(x,y)|\vac\ket\bra\vac|\sigma_{\beta}(0,0)|\vac\ket \n
 & & +\int\d\theta_{1}\d\theta_{2}\bra\vac|\sigma_{\alpha}(x,y)|\theta_{1},\theta_{2}\ket_{+-}^{\phantom{+-}-+}\bra\theta_{2},\theta_{1}|\sigma_{\beta}(0,0)|\vac\ket+\cdots \n
 & = & \bra\vac|\sigma_{\alpha}(x,y)|\vac\ket\bra\vac|\sigma_{\beta}(0)|\vac\ket \n
 & & +\int\d\theta_{1}\d\theta_{2}e^{y(E_{\theta_{1}}+E_{\theta_{2}})-ix(p_{\theta_{1}}+p_{\theta_{2}})} \n 
 & & \times\left(\right.\bra\vac|\sigma_{\alpha}(0,0)|\theta_{1},\theta_{2}\ket_{+-}\bra\vac|\sigma_{\beta}(0,0)|\theta_{2}+i\pi,\theta_{1}+i\pi\ket_{+-}\left.\right)+\cdots \n
 & = & c_{\alpha}c_{\beta}m^{\alpha^{2}+\beta^{2}} \n
 & & -\frac{c_{\alpha}c_{\beta}m^{\alpha^{2}+\beta^{2}}\sin(\pi\alpha)\sin(\pi\beta)}{4\pi^{2}} \n 
 & & \times\left(\int\d\theta_{1}\d\theta_{2}e^{my(E_{\theta_{1}}+E_{\theta_{2}})-ix(p_{\theta_{1}}+p_{\theta_{2}})}\frac{e^{(\theta_{1}-\theta_{2})(\alpha-\beta)}}{\cosh^{2}(\frac{\theta_{1}-\theta_{2}}{2})}\right)+\cdots \label{eq:ffexpansion}
\end{eqnarray}
Since the first term is constant it is the second term which will eventually provide the leading large distance behaviour of $\psi$. First, it is useful to use boost invariance to replace the $(x,y)$ dependence in this term with $r$ dependence, so that the second term in (\ref{eq:ffexpansion}) becomes
\begin{equation}\label{eq:2ndterm}
 -\frac{c_{\alpha}c_{\beta}m^{\alpha^{2}+\beta^{2}}\sin(\pi\alpha)\sin(\pi\beta)}{4\pi^{2}}\int\d\theta_{1}\d\theta_{2}e^{-r(E_{\theta_{1}}+E_{\theta_{2}})}\frac{e^{(\theta_{1}-\theta_{2})(\alpha-\beta)}}{\cosh^{2}(\frac{\theta_{1}-\theta_{2}}{2})}.
\end{equation}
This expression can be further simplified using a change of variables inside the integral. Using the variables
\begin{eqnarray}
 \phi_{1}=\frac{\theta_{1}+\theta_{2}}{2} & \text{and} & \phi_{2}=\frac{\theta_{1}-\theta_{2}}{2}
\end{eqnarray}
we see that the integral in (\ref{eq:2ndterm}) becomes
\begin{eqnarray}
 \frac{1}{2}\int\d\phi_{1}\d\phi_{2}e^{-rm(\cosh(\phi_{1}+\phi_{2})+\cosh(\phi_{1}-\phi_{2})}\frac{e^{2\phi_{2}(\alpha-\beta)}}{\cosh^{2}\phi_{2}} \n
\quad = \frac{1}{2}\int\d\phi_{1}\d\phi_{2}e^{-2rm\cosh(\phi_{1})\cosh(\phi_{2})}\frac{e^{2\phi_{2}(\alpha-\beta)}}{\cosh^{2}\phi_{2}} \n
\quad = \int\d\phi_{2}\frac{e^{2\phi_{2}(\alpha-\beta)}}{\cosh^{2}\phi_{2}}K_{0}(2rm\cosh\phi_{2}) \label{eq:b1}
\end{eqnarray}
where $K_{n}(x)$ is a modified Bessel function.

Armed with this expression we wish to find out what $\Sigma$ looks like at large values of $r$. From the definition (\ref{eq:defSig}) we can write
\begin{eqnarray}
 \Sigma & = & \ln(1-\frac{\sin(\pi\alpha)\sin(\pi\beta)}{4\pi^{2}}\int\d\theta~\frac{e^{2\theta(\alpha-\beta)}}{\cosh^{2}\theta}K_{0}(2mr\cosh\theta))+\cdots \n
 & = & \ln(1-g(r))+\cdots \label{eq:Sigexpansion}
\end{eqnarray}
where $g(r)$ is defined implicitly. For ease of notation we also introduce the constant
\begin{equation}
  \xi=\frac{\sin(\pi\alpha)\sin(\pi\beta)}{4\pi^{2}}.
\end{equation}

To approximate $\psi$ we need to calculate
\begin{equation}
  \label{eq:DSisapprox}
  \left(\pr^{2}+\frac{1}{r}\pr\right)\Sigma
\end{equation}
which quickly becomes very cumbersome when using (\ref{eq:Sigexpansion}). However, at large distances $g(r)$ becomes small so we may expect the Taylor series expansion of the log to still provide a good approximation. Thus we define our approximate solution for $\Sigma$ as
\begin{equation}
  \label{eq:defsigt}
  \St=-g(r)
\end{equation}
and thus
\begin{eqnarray}
  \left(\pr^{2}+\frac{1}{r}\pr\right)\St & = & -(g''(r)+\frac{1}{r}g'(r)) \n
  & = & -4m^{2}\xi\int\d\theta~e^{2\theta(\alpha-\beta)}K_{0}(2mr\cosh\theta).
\end{eqnarray}

We now have a useful approximation to the left hand side of (\ref{eq:myres1}) but to find a usable expression for $\psi$ we must also examine the right hand side. As $r\rightarrow\infty$ (\ref{eq:corrsig}) is expected to tend to a constant, namely $c_{\alpha}c_{\beta}m^{\alpha^{2}+\beta^{2}}$, and so in this limit it must be that $\psi\rightarrow0$ and we may Taylor expand the right hand side of (\ref{eq:myres1}) and only keep the first term:
\begin{equation}
 \frac{m^{2}}{2}\left(1-\cosh(2\psi)\right)=-m^{2}\psi^{2}+\cdots
\end{equation}
Letting our approximate solution be $\psit$ we see that
\begin{equation} \label{eq:psitapp}
 \psit^{2}=4\xi\int\d\theta~e^{2\theta(\alpha-\beta)}K_{0}(2mr\cosh\theta)=J(r)
\end{equation}
where the function $J(r)$ is defined for later convenience. Our goal now is to see how well this approximate solution solves the equation (\ref{eq:myres2}). For comparison we will also examine how well the corresponding equation from \cite{BL97},
\begin{equation}
  \label{eq:BLpsi}
 \D\psi=\frac{m^{2}}{2}\sinh(2\psi)+\frac{4(\alpha-\beta)^{2}}{r^{2}}\tanh(\psi)\left(1-\tanh^{2}(\psi)\right)
\end{equation}
is satisfied $\psit$.

\subsection{Numerical Approximation} \label{sec:numapprox}

To give a first indication as to which solution is correct we will examine how well the approximate solution (\ref{eq:psitapp}) satisfies the equations (\ref{eq:myres2}) and (\ref{eq:BLpsi}) numerically. As the left hand side of both equations is the same it will be sufficient to examine the functions 
\begin{eqnarray}
  (\pr^{2}+\frac{1}{r}\pr)\psit \label{eq:LHSapp} \\
  \frac{m^{2}}{2}\sinh(2\psit)+\frac{4(\alpha-\beta)^{2}}{r^{2}}\tanh\psit(1-\tanh^{2}\psit) \label{eq:RHSBLapp} \\
  \frac{m^{2}}{2}\sinh(2\psit)+\frac{(\alpha-\beta)^{2}}{r^{2}}\tanh\psit(1-\tanh^{2}\psit). \label{eq:RHSmyapp}
\end{eqnarray}

\begin{figure}[ht]
  \centering
  \includegraphics[width=0.9\textwidth]{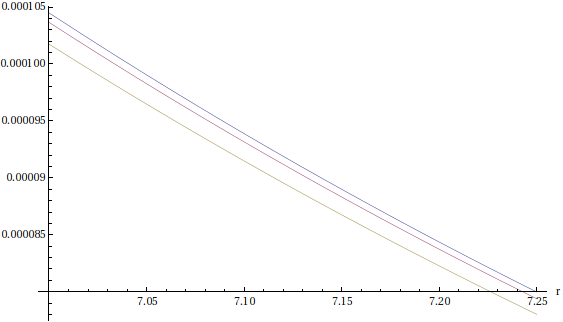}
  \caption{Plot of (\ref{eq:LHSapp}) (blue), (\ref{eq:RHSBLapp}) (yellow) and (\ref{eq:RHSmyapp}) (red).}
  \label{fig:approxplot}
\end{figure}

These functions can be calculated numerically for a fixed value of $(\alpha-\beta)$ and their curves plotted. The value of $(\alpha-\beta)$ must be non-zero as the two equations are equal for $\alpha=\beta$. In figure \ref{fig:approxplot} these curves are plotted with $(\alpha-\beta)=0.3$ and $r\in(7,7.25)$. From this plot it is clear that for modest values of $r$ the expression presented in section \ref{sec:wardidents}, (\ref{eq:RHSmyapp}), is a better approximation of (\ref{eq:LHSapp}) than (\ref{eq:RHSBLapp}). The same behaviour is observed for different values of $(\alpha-\beta)$.

From plots such as figure \ref{fig:approxplot} we see that (\ref{eq:RHSmyapp}) provides a better approximate solution for all values of $r$. At $r=10$, for example (\ref{eq:LHSapp}) is $4.36163\times10^{-6}$, (\ref{eq:RHSmyapp}) is $4.34464\times10^{-6}$ while (\ref{eq:RHSBLapp}) is $4.30542\times10^{-6}$. So in this case and using the equation presented in this paper the approximate solution $\psit$ is out by $0.39\%$. When the equation (\ref{eq:RHSBLapp}) is used this difference jumps to $1.39\%$. Again the same pattern appears when different values of $(\alpha-\beta)$ and $r$ are used.

The numerical evidence presented in this subsection strongly suggests that the equation presented in this paper is indeed the correct parametrisation for the correlation functions twist fields.


\subsection{Analytic Approximation} \label{sec:anapprox}

With the strong numerical evidence pointing to (\ref{eq:myres2}) being the correct expression for the function parametrising the correlation functions of twist fields we will now look for some analytic evidence to support this. The objective is to expand both sides of (\ref{eq:myres2}) and test whether or not the leading $r$ terms cancel when the approximation $\psi=\psit$ is applied.

To begin with, as we know that $\psi$ is small at large values of $r$, we may expand the right hand side of (\ref{eq:myres2}) and only keep the linear term:
\begin{equation}
  \label{eq:RHSlinexpansion}
  \frac{m^{2}}{2}\sinh(2\psi)+\frac{(\alpha-\beta)^{2}}{r^{2}}\tanh\psi(1-\tanh^{2}\psi) = \left(m^{2}+\frac{(\alpha-\beta)^{2}}{r^{2}}\right)\psi+\cdots 
\end{equation}
We now want to insert our approximation of $\psi$ and examine the equation
\begin{equation}
  (\pr+\frac{1}{r}\pr)\psit=\left(m^{2}+\frac{(\alpha-\beta)^{2}}{r^{2}}\right)\psit+\cdots
\end{equation}
It is convenient at this point to use the function $J(r)$ defined in (\ref{eq:psitapp}) so that this equation becomes, neglecting higher order terms,
\begin{equation}
  \label{eq:Jrel}
  \frac{J''J-\frac{1}{2}(J')^{2}+\frac{1}{r}J'J}{2J^{3/2}}\simeq\left(m^{2}+\frac{(\alpha-\beta)^{2}}{r^{2}}\right)\sqrt{J}
\end{equation}
and to further ease our calculations this expression can be rearranged to give
\begin{equation} \label{eq:Jlineq1}
 J''J-\frac{1}{2}(J')^{2}+\frac{1}{r}J'J\simeq2\left(m^{2}+\frac{(\alpha-\beta)^{2}}{r^{2}}\right)J^{2}.
\end{equation}

In order to progress further we need:
\begin{eqnarray}
 J(r) & = & 4\xi\int\d\theta~e^{2\theta(\alpha-\beta)}K_{0}(2mr\cosh\theta) \\
 J'(r) & = & -8m\xi\int\d\theta~e^{2(\alpha-\beta)\theta}\cosh\theta K_{1}(2mr\cosh\theta) \\
 J''(r) & = & 16m^{2}\xi\int\d\theta~e^{2(\alpha-\beta)\theta}\cosh^{2}\theta K_{0}(2mr\cosh\theta) \n
 & & +8m\xi\int\d\theta~e^{2(\alpha-\beta)\theta}\frac{\cosh\theta}{r}K_{1}(2mr\cosh\theta).
\end{eqnarray}
From this we see that the second term in the $J''J$ term of (\ref{eq:Jlineq1}) cancels with the $J'J/r$ term so there are only two terms left to deal with.

Next we note that there are no $(\alpha-\beta)^{2}$ terms in these derivatives so we need to eliminate these from (\ref{eq:Jlineq1}). This is done by integrating $J(r)$ by parts, integrating the exponential term to bring down $1/(\alpha-\beta)$ terms:
\begin{eqnarray}
 \int\d\theta~e^{2(\alpha-\beta)\theta}K_{0}(2mr\cosh\theta) & = & -\frac{mr}{\alpha-\beta}\int\d\theta~e^{2(\alpha-\beta)\theta}K_{1}(2mr\cosh\theta)\sinh\theta \n
 & = & \frac{mr}{2(\alpha-\beta)^{2}}\int\d\theta~e^{2(\alpha-\beta)\theta}\left(2mrK_{0}(2mr\cosh\theta)\sinh^{2}\theta\right. \n & & \left.+\frac{\sinh^{2}\theta}{\cosh\theta}K_{1}(2mr\cosh\theta)-K_{1}(2mr\cosh\theta)\cosh\theta\right) \n
 & = & \frac{m^{2}r^{2}}{(\alpha-\beta)^{2}}\int\d\theta~e^{2(\alpha-\beta)\theta}K_{0}(2mr\cosh\theta)(\cosh^{2}\theta-1) \n
 & & -\frac{mr}{2(\alpha-\beta)^{2}}\int\d\theta~e^{2(\alpha-\beta)\theta}\frac{1}{\cosh\theta}K_{1}(2mr\cosh\theta). \label{eq:Jbyparts}
\end{eqnarray}

Now writing the right hand side of (\ref{eq:Jlineq1}) as 
\begin{equation}
 m^{2}J^{2}+\frac{(\alpha-\beta)^{2}}{r^{2}}J^{2}
\end{equation}
and using the above integration by parts to rewrite one of the $J$'s multiplying the $(\alpha-\beta)^{2}$ term we see that the $m^{2}$ term is cancelled by part of the first integral in (\ref{eq:Jbyparts}) and the other part of this integral cancels some of the integral from the $J''J$ term on the left hand side. So bringing all the remaining terms of (\ref{eq:Jlineq1}) together we have
\begin{multline} \label{eq:collectedterms}
 32m^{2}\xi^{2}\left(\int\d\theta~e^{2(\alpha-\beta)\theta} K_{0}(2mr\cosh\theta)\right) \\ \times\left(\int\d\theta~e^{2(\alpha-\beta)\theta}(\frac{1}{2mr\cosh\theta}K_{1}(2mr\cosh\theta)+\cosh^{2}\theta K_{0}(2mr\cosh\theta))\right) \\
-32m^{2}\xi^{2}\left(\int\d\theta~e^{2(\alpha-\beta)\theta}\cosh\theta K_{1}(2mr\cosh\theta)\right)^{2}\simeq0
\end{multline}
from which we want to extract the leading large $r$ behaviour. Expanding the Bessel functions for large $r$ gives leading terms proportional to $1/\sqrt{r}$:
\begin{multline} \label{eq:leadingterms}
 32m^{2}\xi^{2}\left(\int\d\theta~e^{2(\alpha-\beta)\theta}\sqrt{\frac{\pi}{4mr\cosh\theta}}e^{-2mr\cosh\theta}\right) \\ \times \left(\int\d\theta~e^{2(\alpha-\beta)\theta}\cosh^{2}\theta \sqrt{\frac{\pi}{4mr\cosh\theta}}e^{-2mr\cosh\theta}\right) \\
-32m^{2}\xi^{2}\left(\int\d\theta~e^{2(\alpha-\beta)\theta}\cosh\theta \sqrt{\frac{\pi}{4mr\cosh\theta}}e^{-2mr\cosh\theta}\right)^{2}.
\end{multline}
While these terms do not cancel directly we observe that the main contribution from each integrand is from the region around $\theta=0$ and so the first order saddle point approximation of these integrals does indeed vanish, as we had hoped.


\section{Summary}\label{sec:sum}

As stated in the introduction we have derived the equations parametrising the correlation function
\begin{equation}
  \bra\sigma_{\alpha}(x,y)\sigma_{\beta}(0,0)\ket
\end{equation}
following the method of \cite{DS11}. We have also presented strong evidence that the equations derived in this paper provide the correct parametrisation of the correlation function. 

With this derivation we have retrieved the known results regarding twist fields in the Dirac model using a direct and straight forward method. Unlike the equations of \cite{DS11} the equations found here require rotation symmetry and so the result is not as strong. One extension of this work would be to determine if it would be possible to derive similar differential equations where the states considered have more general symmetries.

The method presented has already been applied to Ising field theory \cite{FZ03} but its success in this more general setting suggests that it may have wider applications. It would be interesting to discover if this method could be successfully applied to other massive field theories to produce new results.


\paragraph*{Acknowledgements}
The author wishes to thank Benjamin Doyon for his continued support and guidance and Ed Corrigan and Peter Bowcock for their help in preparing this manuscript. This work was conducted under an EPSRC studentship.


\end{document}